\documentclass[a4paper, oneside, 12pt]{scrartcl}

\usepackage[utf8x]{inputenc}
\usepackage[T1]{fontenc}
\usepackage[english]{babel}

\makeatletter
\DeclareOldFontCommand{\rm}{\normalfont\rmfamily}{\mathrm}
\DeclareOldFontCommand{\sf}{\normalfont\sffamily}{\mathsf}
\DeclareOldFontCommand{\tt}{\normalfont\ttfamily}{\mathtt}
\DeclareOldFontCommand{\bf}{\normalfont\bfseries}{\mathbf}
\DeclareOldFontCommand{\it}{\normalfont\itshape}{\mathit}
\DeclareOldFontCommand{\sl}{\normalfont\slshape}{\@nomath\sl}
\DeclareOldFontCommand{\sc}{\normalfont\scshape}{\@nomath\sc}
\makeatother

\usepackage[a4paper,  centering, bindingoffset=0mm, inner=25mm, outer=25mm, top=26mm, bottom=36mm, heightrounded]{geometry}

\usepackage{graphicx}
\usepackage{float}
\usepackage{cite}

\usepackage{amsmath}	
\usepackage{amssymb}
\usepackage{amsfonts}
\usepackage{mathrsfs}
\usepackage{mathtools}
\usepackage{amsthm}
\usepackage{bm}
\usepackage{esint}     

\usepackage{braket}
\usepackage{slashed}

\usepackage[]{todonotes}

\usepackage[font=small,labelfont=bf, width=.95\textwidth]{caption} 

\usepackage{verbatim}
\usepackage{textcomp}
\usepackage{enumitem}
\usepackage{circuitikz}

\usepackage[toc]{appendix}

\usepackage{layout}
\usepackage{microtype}
\usepackage{bookmark}
\usepackage{color}
\usepackage[]{xcolor}

\newcommand{\dd}{\mathrm{d}}

\newcommand{\ii}{\mathrm{i}}

\makeatletter
\newcommand{\subalign}[1]{%
	\vcenter{%
		\Let@ \restore@math@cr \default@tag
		\baselineskip\fontdimen10 \scriptfont\tw@
		\advance\baselineskip\fontdimen12 \scriptfont\tw@
		\lineskip\thr@@\fontdimen8 \scriptfont\thr@@
		\lineskiplimit\lineskip
		\ialign{\hfil$\m@th\scriptstyle##$&$\m@th\scriptstyle{}##$\crcr
			#1\crcr
		}%
	}
}
\makeatother
\usepackage{tikz}

\usetikzlibrary{decorations.pathmorphing}
\usetikzlibrary{shapes.geometric}
\usetikzlibrary{matrix, calc, arrows}
\tikzset{snake it/.style={decorate, decoration=snake}}
\usetikzlibrary{decorations.markings}

\tikzset{%
	dots/.style args={#1per #2}{%
		line cap=round,
		dash pattern=on 0 off #2/#1
	}
}

\definecolor{unamblue}{rgb}{0.0, 0.0, 0.0}

\usepackage[]{hyperref}
\hypersetup{
	colorlinks=true,%
	citecolor=unamblue,%
	filecolor=unamblue,%
	linkcolor=unamblue,%
	urlcolor=unamblue,
	bookmarksnumbered=true,     
	bookmarksopen=true,         
	bookmarksopenlevel=1,       
	pdfstartview=Fit,           
	pdfpagemode=UseOutlines,
	pdfpagelayout=TwoPageRight
}

\usepackage{graphicx}
\usepackage{tikz}
\usepackage{slashed}
\usepackage{epstopdf}
\usepackage{verbatim}	
\usepackage{blkarray}
\usepackage{ytableau}

\usepackage[]{todonotes}
\usepackage{subcaption}
\usepackage{verbatim}
\usepackage{textcomp}
\usepackage{enumitem}

\usepackage[toc]{appendix}

\usepackage{layout}
\usepackage{microtype}
\usepackage{bookmark}
\usepackage{color}
\usepackage[]{xcolor}

\makeatother
\usepackage{tikz}
\usetikzlibrary{decorations.pathmorphing}
\usetikzlibrary{shapes.geometric}
\usetikzlibrary{matrix, calc, arrows}
\tikzset{snake it/.style={decorate, decoration=snake}}
\usetikzlibrary{decorations.markings}

\tikzset{%
	dots/.style args={#1per #2}{%
		line cap=round,
		dash pattern=on 0 off #2/#1
	}
}

\definecolor{unamblue}{cmyk}{1 0.79 0.12 0.59}

\usepackage[]{hyperref}
\hypersetup{
	colorlinks=true,%
	citecolor=unamblue,%
	filecolor=unamblue,%
	linkcolor=unamblue,%
	urlcolor=unamblue,
	bookmarksnumbered=true,     
	bookmarksopen=true,         
	bookmarksopenlevel=1,       
	pdfstartview=Fit,           
	pdfpagemode=UseOutlines,
	pdfpagelayout=TwoPageRight
}

\usepackage{graphicx}
\usepackage{tikz}
\usepackage{slashed}
\usepackage{epstopdf}
\usepackage{verbatim}	
\usepackage{blkarray}
\usepackage{ytableau}
\usepackage{eufrak}

\newcommand{\lotop}[3]
{\begin{tikzpicture}[thick]
	\draw[draw, snake it] (-0.5,0)--(-0.5,-1)node[below]
	{#1};
	\draw[<-,,<-=stealth, draw] (-0.8,-0.7)--(-0.8,-0.2) node[below left]
	{$k_1$};
	\draw[draw, snake it] (0.5,0)--(0.5,-1)node[below]
	{#2};
	\draw[<-,,<-=stealth, draw] (0.8,-0.7)--(0.8,-0.2) node[below right]
	{$k_2$};
	\path [ draw] (0.5,0) --(1,0);
	\path [draw, #3]    (-0.5,0) --(.5,0);
	\path [draw]    (-1,0) --(-0.5,0);
	\coordinate (A) at (-0.5,0);   \filldraw (A) circle (2.pt);
	\coordinate (B) at (0.5,0);   \filldraw (B) circle (2.pt);
	\end{tikzpicture}
}
\newcommand{\lotopB}[3]
{
\begin{tikzpicture}[thick]
\draw[draw, snake it] (-0.5,0)--(0.5,-1)node[below]
{#1};
\draw[<-,,<-=stealth, draw] (-0.8,-0.7)--(-0.8,-0.2) node[below left]
{$k_1$};
\draw[draw, snake it] (0.5,0)--(-0.5,-1)node[below]
{#2};
\draw[<-,,<-=stealth, draw] (0.8,-0.7)--(0.8,-0.2) node[below right]
{$k_2$};
\path [ draw] (0.5,0) --(1,0);
\path [draw, #3]    (-0.5,0) --(.5,0);
\path [draw]    (-1,0) --(-0.5,0);
\coordinate (A) at (-0.5,0);   \filldraw (A) circle (2.pt);
\coordinate (B) at (0.5,0);   \filldraw (B) circle (2.pt);
\end{tikzpicture}}

\newcommand{\lotopC}[3]
{
\begin{tikzpicture}[thick]
\draw[draw, snake it] (0,0)--(-0.5,-1)node[below]
{#1};
\draw[<-,,<-=stealth, draw] (-0.8,-0.7)--(-0.8,-0.2) node[below left]
{$k_1$};
\draw[draw, snake it] (0,0)--(0.5,-1)node[below]
{#2};
\draw[<-,,<-=stealth, draw] (0.8,-0.7)--(0.8,-0.2) node[below right]
{$k_2$};
\path [ draw] (0.5,0) --(1,0);
\path [draw, #3]    (-0.5,0) --(.5,0);
\path [draw]    (-1,0) --(-0.5,0);
\coordinate (A) at (0,0);   \filldraw (A) circle (2.pt);
\end{tikzpicture}}

\newcommand{\nlotop}[3]{ \begin{tikzpicture}[thick]
	\draw[draw, snake it] (-2,0)--(-2,-1)node[below ]
	{#1};
	\draw[<-,,<-=stealth, draw] (-2.2,-0.7)--(-2.2,-0.2) node[ below left]
	{$k_1$};
	\draw[draw, snake it] (-1,0)--(-1,-1)node[below]
	{#2};
	\draw[<-,,<-=stealth, draw] (-1.2,-0.7)--(-1.2,-0.2) node[ below left]
	{$k_2$};
	\draw[draw, snake it] (0,0)--(0,-1)node[below]
	{#3};
	\draw[<-,,<-=stealth, draw] (-0.2,-0.7)--(-0.2,-0.2) node[ below left]
	{$k_3$};
	\path [draw]
	(-2.5,0) --(-1,0);
	\path [draw]
	(-1,0) --(0,0);
	\path [draw]
	(0,0) --(0.5,0);
	\path [draw]
	(-2,0) --(-1,0);
	\coordinate (A) at (-1,0);   \filldraw (A) circle (2.pt);
	\coordinate (B) at (-2,0);   \filldraw (B) circle (2.pt);
	\coordinate (C) at (0,0);   \filldraw (C) circle (2.pt);
	\end{tikzpicture}}

	{}

\title{\Huge On kinetic theories with color and spin  from amplitudes 
	\\}

\author{ \normalfont\normalsize  Leonardo de la Cruz\\[2mm]
	\emph{\normalfont\small \em Dipartimento di Fisica e Astronomia ``Augusto Righi'', Universit\`a di Bologna}\\
	\emph{\normalfont\small \em and INFN Sezione di Bologna, via Irnerio 46, I-40126 Bologna, Italy}	
}
\date{%
		$\,$%
	\\[2\baselineskip]
	\normalfont\normalsize%
	\parbox{0.8\linewidth}{%
		{\bf \sf Abstract}. 
We extend a previously developed approach to relate thermal currents in the  high temperature regime and classical limits of amplitudes. We  consider the bi-adjoint scalar theory, which has the basic structure of a cubic theory and which is related to QCD and  gravity through the double copy. In addition, we consider a generalization of scalar QED to model classical spin, where massive scalars are complex  higher-spin fields.  We derive Vlasov-type kinetic equations   for  bi-adjoint scalars and study their iterative solutions, while for QED we use well-known kinetic equations. In both cases we find consistency between these solutions and the amplitude-like approach.
  
	}
}

\begin{document}

\maketitle
\thispagestyle{empty}
\newpage
\tableofcontents

\section{Introduction} 
Kinetic theory 
is useful to describe systems out of equilibrium for a wide range of many-particle systems at classical and quantum
level. Its quantum version is based on Wigner functions,  which are quantum analogs of classical distribution functions\footnote{Quantum kinetic theory has been thoroughly reviewed in Ref.\cite{Hidaka:2022dmn}.}. 
At quantum level they appear
from the Wigner transform   of the density matrix operator $\hat \rho(t)$, which for a single particle reads \cite{Wigner:1932eb}
\begin{align}
W(x, p; t):=\int \dd u \  e^{-\ii p u}
\braket{x+\frac{1}{2}u| \hat \rho(t)|
	x-\frac{1}{2}u},
\end{align}
where $x, p$ are phase-space variables. 
A suitable classical limit renders it as the classical distribution function $f(x, p )$. In the collision-less case the distribution function $f(x, p)$ satisfies 
Liouville's equation
$\frac{\dd f}{\dd \tau}=0$,
which gives rise to the Boltzmann equation.
The Wigner transform 
can be generalized to describe quantum fields. Then, from QFT it is possible to deduce an equation for the Wigner function which satisfies classical kinetic equations thus confirming the analogy between Wigner functions and classical phase-space distributions. 

Another alternative is to maintain the point-particle picture and bring together Wigner transformations and the Schwinger-Keldysh formalism in a 
first-quantized worldline approach\footnote{The worldline approach has been reviewed in Ref.\cite{Schubert:2001he}}, thus allowing the construction of phase-space distributions and a derivation of classical kinetic equations of Vlasov type, which may include color and spin degrees of freedom \cite{Mueller:2019gjj}. Following the Schwinger-Keldysh strategy, the action  expressed in terms of retarded and advanced fields contains boundary terms that one can formally relate to the Wigner transform of the density operator, which is  typical in finite temperature QFT \cite{Jeon:2004dh, Jeon:2013zga}.
Also typical in this context is that
the path integral over advanced variables imposes classical equations of motion--- which is also the conclusion one obtains in the worldline approach--- and hence the remaining path integrals are  over retarded variables. 

In Ref.~\cite{delaCruz:2020cpc},  iterative solutions of
kinetic equations were mapped  to certain off-shell currents in the classical limit, understood as the limit $\hbar \to 0$. There, we applied the  Kosower-Maybee-O'Connell (KMOC) \cite{Kosower:2018adc} formalism to deal with off-shell currents rather than amplitudes in the forward limit. KMOC  was originally proposed in the context of ongoing
efforts to extract classical information from scattering amplitudes (see Refs.\cite{Bjerrum-Bohr:2022blt,Kosower:2022yvp} and references therein for reviews). These efforts are concerned with matching solutions of classical equations of motion and scattering amplitudes, which is  similar to the scenario we  discuss here,  the additional ingredient being that equations of motion are now coupled to the Boltzmann equation for the distribution function.
These solutions were constructed in  the approximation where  the distribution function $f$ can be perturbatively expanded in powers of the coupling constant. This situation arises e.g., in the description of the so-called hard thermal loops\cite{Braaten:1989mz,  Frenkel:1989br, Braaten:1990az, Taylor:1990ia,
	Frenkel:1991ts}, which are relevant in the high temperature limit of QCD and  can be mapped to solutions of kinetic equations \cite{Kelly:1994ig}.

In the context of scattering amplitudes \cite{Elvang:2013cua, Weinzierl:2016bus}, bi-adjoint scalars are important ingredients to understand the relationship between QCD and gravity through the so called "double copy", which has been reviewed in  Ref.\cite{Bern:2019prr}. At tree-level closed formulas for massless and massive amplitudes are available from the Cachazo-He-Yuan representation \cite{Cachazo:2013hca}. Bi-adjoint scalars  have been explored from many angles  ranging from  more mathematically-inspired ones \cite{Cachazo:2022vuo,Cachazo:2019apa}  to studies of its basic properties such as renormalization
\cite{Gracey:2020baa,Gracey:2020tkk} or its exact solutions \cite{White:2016jzc,Bahjat-Abbas:2018vgo} to name just a few.  Double-copy-type  relations also exist classically 
\cite{Monteiro:2014cda} and, more importantly, they   also exist  for  particles carrying spin \cite{Goldberger:2017ogt}. Since  bi-adjoint scalar amplitudes encode the kinematic structure of QCD and gravity it raises the question of whether this is also the case at finite temperature even for particles carrying spin.

In this paper, we  will extend the approach proposed in Ref.\cite{delaCruz:2020cpc} to include bi-adjoint scalars to start exploring the above questions. In addition, we will consider classical spin from amplitudes which we will model through a generalization of scalar electrodynamics in which the scalar fields are  higher-integer-spin fields\cite{Bern:2020buy}. As we will see, the amplitudes-based approach  leads to consistent results that match perturbative solutions of their kinetic equations.

The remainder of the paper is organized as follows. In Section \ref{color-less}, we review semi-classical kinetic theory with spin and color and outline the amplitude-like approach to thermal currents.
In Section \ref{Bi-adjoint}, we introduce color in the context of the bi-adjoint scalar model. In Section  \ref{gauge-spin}, we consider classical spin. Our conclusions are presented in Section \ref{conclusions}. 

\section{Amplitudes approach to thermal currents}
\label{color-less}

\subsection{Thermal currents and the classical limit}
Let us consider a  distribution function $f(x,p,c,s)$, which in addition to the coordinate $x^\mu$ and momentum $p^\mu$ depends on two continuous variables $c^a$ and $s^\mu$, describing color and spin degrees of freedom, respectively. The Liouville's equation in the collision-less case can be expressed as \cite{Heinz:1984yq}
\begin{align}
\frac{\dd}{\dd  \tau} f(x,p,c,s) =
\left(\dot x^\mu 
\frac{\partial }{\partial x^\mu}+ \dot p^\mu 
\frac{\partial }{\partial p^\mu}+\dot c^a\frac {\partial}{\partial c^a}+\dot s^\mu \frac{\partial}{\partial s^\mu} \right) f(x,p,c,s) =0,
\end{align}
where the Vlasov-Boltzmann equation can be obtained from the above after inserting the equations  of motion  of the phase-space variables, which are generalized versions of Wong equations \cite{Wong:1970fu}. In the color-less case the spin vector satisfies the BMT equation 
\cite{Bargmann:1959gz}.
The associated color currents of the particles are obtained from
\begin{align}
J_a^\mu(x)= g\int\dd\Phi (p) \,\int\dd c \,\int\dd s\, c_a\, (p^\mu +S^{\mu\nu}\partial_\nu) f(x,p,c,s),
\label{current-1}
\end{align}
where $S^{\mu\nu}$ is the spin tensor and  $\dd\Phi(p)$ is the usual Lorentz Invariant Phase space. The invariant measures for color and spin space are 
$\dd c$ and $\dd s$,  respectively. The invariant measures are given by
\cite{Elze:1989un}
\begin{align}
\dd \Phi (p):=& \frac{\dd ^4 p}{(2\pi)^3}
\, \Theta(p_0) \delta(p^2-m^2),\\
\dd c:=&  \dd^8 c\, c_R \delta(c^a c^b \delta^{ab}-q_2 ) \delta(d^{abc}c^a c^bc^c-q_3 ), \label{DIPSC} \\
\dd s:=& \dd^4 s\, c_S \delta(s^\mu s_\mu+2\mathfrak 	s^2) \delta(p^\mu s_\mu),  
\end{align} %
where $q_2$, $q_3$, $\mathfrak s^2$ are Casimir invariants\footnote{We assume that classical kinetic theory to be valid in the approximation where color and spin representations are large so that color factors and spin vectors commute. }. The factors $c_S$ and $c_R$ ensure that spin and color measures are normalized to unity. We have set the gauge group to be $SU(3)$ and for spin to be $SU(2)$.  The integration over phase space can be done using the relation between the spin vector and spin tensor
\begin{align}
S^{\mu\nu}=-\frac 1 {m} \epsilon^{\mu\nu\rho \sigma}p_\rho s_\sigma,
\label{spin-tensor-to-vector}
\end{align}
where $\epsilon^{0123}=1$.    There is a one to one correspondence between  classical color factors $c^a$ and the generators of the gauge group and similarly for spin. Therefore the invariants $q_2$, $q_3$ and $\mathfrak s^2$	are defined through the traces of the generators of the gauge group
\begin{align}
\text{Tr}(T^a T^b)= C_2 \delta^{a b},
\label{basic-trace}
\end{align}
where for $SU(N)$ the quadratic Casimir $C_2=N$ in the adjoint representation and $C_2=1/2$ in the fundamental.   Thus, properties of integration over classical phase spaces are equivalent to traces of color-factors, namely 
\begin{equation}
\begin{aligned}
\int \dd c\, c^a =& \int \dd s\, s^\mu =0,\\
\int \dd  c\,  c^a c^b=&  C_2 \delta^{ab},\\
\int \dd  s\, \mathit s^\mu \mathit s^\nu=& - \frac {2} 3 \mathfrak s^2 \left(\eta^{\mu\nu}-\frac{p^\mu p^\nu}{m^2} \right).
\end{aligned}
\label{identities-phase-space}
\end{equation}
The last  equalities express the fact that $c^2$ and $s^2$ are constants of the motion. For spin, the last equality is equivalent to the trace  of the spin tensor and $\mathfrak s^2$ is proportional to the eigenvalue of the Casimir operator \cite{Bhadury:2020puc,Weickgenannt:2020aaf}. More generally, the projectors in parenthesis will depend on the matter content \cite{Hidaka:2022dmn}.

We may evaluate the currents of Eq.\eqref{current-1} iteratively by assuming a perturbative expansion of $f(x,p,c,s)$ in powers of the coupling constant $g$ around the equilibrium state $f(p_0)$, where the distribution depends only on the energy $p_0$ (see Ref.\cite{Litim:2001db} for review of this approach). We allow the equilibrium distribution to be either Bose-Einstein ($f_ -(p_0)$)  or Fermi-Dirac  ($f_+(p_0)$)  so strictly speaking this approach is semi-classical.  For QCD
the relation between color currents and thermal currents in the high temperature approximation (hard thermal loops) is conveyed by  
\begin{align}
J^a_\mu(x) = \Pi^{ab}_{\mu\nu}A^{b\nu} 
+ \frac 1 2\Pi^{abc}_{\mu\nu\rho}A^{b\nu}A^{c\rho} +\dots,
\end{align}
where the currents $\Pi^{a_1 \cdots a_n}_{\mu_1 \cdots \mu_n}$ thus obtained  match those obtained
in the high temperature limit of QCD\cite{Kelly:1994ig, Kelly:1994dh}.
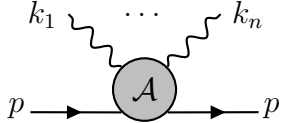
\begin{figure}
	\centering
\begin{tikzpicture}[thick, transform shape]
\node[circle, fill=lightgray, draw] (c) at (0,0){$\mathcal A$};
\draw[draw, snake it] (c.north west)--(-1.,1)node[left]
{$\qquad k_1$};
\draw[draw, snake it] (c.north east)--(1.,1)node[right]
{$k_n$};
\draw[draw] (c.south east)--(1.5,-0.30) node[right,     currarrow,
pos=0.5, 
xscale=1,
sloped,
scale=1]
{$\quad \ \, p$};
\draw[draw] (c.south west)--(-1.5,-0.30)node[left,    currarrow,
pos=0.5, 
xscale=1,
sloped,
scale=1]
{$\mkern-36mu p$};
\node(A) at (0,1){$\dots$};
\end{tikzpicture}
	\caption{Regulated off-shell current in the forward limit.  The blob represents a sum over tree-level Feynman diagrams where diagrams generating a zero-momentum internal edge are suppressed.  Wavy lines represent off-shell outgoing soft particles.}
	\label{current-off-shell}
\end{figure}

In Ref.\cite{delaCruz:2020cpc}, we proposed to  obtain thermal currents by taking the classical limit of a regulated off-shell current in the forward limit adapting the KMOC formalism. Let us briefly recap the main points. The forward limit is, in general, singular so a regularization scheme is required. Here, it  is understood as discarding  diagrams with a zero momentum internal edge \cite{Runkel:2019zbm}. Let
$\mathcal S$ be the set of those diagrams and let $\mathcal F$ be the set of all diagrams contributing to the amplitude, then  the regularized current is defined by
\begin{align}	
\mathcal A^n (p, k_1, ,\dots, k_n,p):=\sum_{G\in \mathcal F   \backslash \mathcal S}
d(G),
\end{align}
where $d(G)$ is a rational  expression of the form $N(G)/D(G)$. Denoting by $k:=(k_1, \dots, k_n)$ the $n$-tuple of external off-shell momenta and, with the understanding that the amplitude is computed in the forward limit, we sometimes write $\mathcal A^n(p,k)$. The classical limit is obtained as follows: Compute the
Laurent expansion in powers of $\hbar$ of the $n+2$ current $\mathcal A^n (p, k_1, \dots, k_n, p)$ where the momenta of the particles $1, \dots, n $ is considered soft and off-shell  ($k_i^2\ne 0$), while  massive particles carrying momenta $p$ are on-shell ($p^2=m^2$). Soft momenta scale with $\hbar$, i.e,  $k\to \hbar k$, implementing the distinction between  momentum and  wavenumber of a particle, which is an important step in the KMOC formalism. 
In practice
the current can be computed e.g., from Feynman graphs (See Fig.\ref{current-off-shell}). 
Recall that we are using units in which $k_B=c=1$ but keeping $\hbar \ne 1$ since we are interested in the classical limit. 
We adopt the convention that our classical results  depend on the 
dimensionless coupling $g=\bar{g} \sqrt{\hbar}$ and $e=\bar{e} /\sqrt{\hbar}$, and that the external momenta is associated to wavenumbers. Suppressing	 color and Lorentz indices 
the classical limit of the current is given by 
\begin{align}
\bar {\mathcal A}^n (p,k)= \widehat{\text{Tr}}\left(\lim_{\hbar \to 0}
\mathcal A^n (p, \hbar k)\right), \end{align}
where the trace  is defined by
\begin{align}
\widehat{\text{Tr}} (\bullet) := \begin{cases} 	
\hbar^{n-2} \text{Tr} (\bullet) & \text{QCD}, \\
\text{Id} (\bullet) & \text{QED} \ \text{and} \ \text{gravity},
\end{cases}
\label{trace-tilde-def}
\end{align}
where $\text{Id}$ is the identity operator and the $\hbar^{n-2}$ is required on dimensional grounds. We then have a simple relation between thermal currents and the classical limit\footnote{Our conventions for the Fourier transform are
	$g(x)=\int  \dd ^d k/(2\pi)^d \, e^{-\ii k\cdot x} \tilde g (k)
	$
} 
\begin{align}
\Pi^n (k)= \int \dd \Phi (p)  \,f(p_0) \ \bar{\mathcal A}^n (p, k)
\label{off-shell-current}
\end{align}
where $f(p_0)$ is the distribution function at equilibrium.  If we are interested in fermions we will define the distribution function with a minus sign due to the presence of a fermion loop. This relation represents a map between the classical limit and the high temperature in the forward limit\cite{Frenkel:1991ts}.  In Sections \ref{Bi-adjoint} and \ref{gauge-spin} we will add two more cases where Eq.\eqref{off-shell-current} holds.

Let us briefly
mention that color factors in calculations are defined through 
\begin{align}
\braket{p_i| \mathbb C^a|p_j}:= (C^a)_i^{\,j}=\hbar (T^a)_{i}^{\, j}, \label{color-expectation}
\end{align}
where $\mathbb C^a$ are operators that realize the Lie algebra of the gauge group
\begin{align}
[\mathbb C^a, \mathbb C^b]=\ii  \hbar f^{abc}\mathbb C^c.
\end{align}
Color charge operators can be explicitly derived from the Noether procedure (see Sections 2 and 4 of Ref.\cite{delaCruz:2020bbn}). 
Classical color charges $c^a:=\braket{\psi|\mathbb C^a|\psi}$ are obtained as expectation values of those operators taken from appropriate coherent states $\ket{\psi}$. Using these estates, color charge operators satisfy the important property (shown in Appendix A of  Ref.\cite{delaCruz:2020bbn}) of factorization
\begin{align}
\braket{\psi|\mathbb C^a \mathbb C^b|\psi}=\braket{\psi|\mathbb C^a|\psi}
\braket{\psi|\mathbb C^b|\psi}+\mathcal O (\hbar),
\label{color-factorization}
\end{align} 
so classical color charges commute. 
\section{Color: the bi-adjoint scalar}\label{Bi-adjoint}
The field theory Lagrangian of the model is given by \cite{Luna:2015paa,White:2016jzc}
\begin{align} 
\mathcal L_{\text {BA}}:={\cal L _{\text {BA}}}( \varphi, \partial \varphi ) = \frac12 \partial_\mu \varphi^{a \alpha} \partial^\mu \varphi^{a \alpha} 
-\frac{m^2}{2} \varphi^{a \alpha}\varphi^{a \alpha} 
+\frac{y}{3!} f^{abc} \tilde f^{\alpha \beta \gamma}
\varphi^{a \alpha} \varphi^{b \beta} \varphi^{c \gamma} \:, 
\label{lag-phi3}
\end{align}
where $m$ is the mass and $y$ the coupling constant. The bi-adjoint scalar field  $\varphi^{a\alpha}$ transforms under the adjoint representation for each factor of its globally symmetry group $G \times \tilde G$. The Lie algebra for each factor has the form $[T^a, T^b]=\ii f^{abc} T^c$ and the adjoint representation is given by its structure constants, i.e.
$(T_A^a)^{b}_{\ c}=- \ii f^{abc}$. Throughout we  use Greek indices   for the group $\tilde G$. Before developing a kinetic theory for bi-adjoint scalars let us briefly discuss how these are computed from the KMOC formalism. 

\subsection{Off-shell currents for bi-adjoint scalars}

We consider the theory in $d=6-2\epsilon$  dimensions, where the bi-adjoint scalar can be renormalized with a dimension-less coupling. We will keep the same definition of color factors in the adjoint
representation as in Eq.\eqref{color-expectation}
so color factors have dimensions of $\hbar$
and, as it should be, they are  absent in the classical limit. Since the action has units\footnote{$M$ and $L$ denote mass and length, respectively.} of $ML$,   in $d=6$ the field
$\varphi^{a\alpha}$ has dimensions $\sqrt{M/L^3}$ so the dimensionful coupling $\bar y$ has units of $1/\sqrt{ML}$, and therefore  the dimensionless coupling scales as $\sqrt{\hbar} \bar y$. Then,  thermal currents are obtained through Eq.\eqref{off-shell-current} with the trace $\widehat{\text{Tr}} (\bullet)$ now defined as:
\begin{align}
\widehat{\text{Tr}} (\bullet):=\text{Tr} (\bullet) \
\widetilde{\text{Tr}} (\bullet),
\label{trace-bi-adjoint}
\end{align}
where it is understood that first and second traces are referred to the untilded and tilded color factors, respectively. This adds another instance of Eq.\eqref{trace-tilde-def}. Notice however that for a general representation $R$
\cite{vanRitbergen:1998pn}
\begin{align}
\text{Tr} (T_R^{a} T_R^{b}T_R^{c})=d_R^{abc}+ \frac \ii 2 f^{abc} I_2 (R),
\end{align}
where $I_2(R)$ is the index of the representation. Since in the adjoint representation $d_R^{abc}=0$, classical phase space integration is reproduced through the replacement $f^{abc} \to d^{abc}$ at the end of the calculation (see Appendix \ref{3-point-bi-adjoint}).  
\begin{figure}[t]
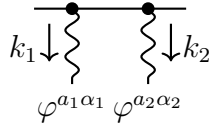

	\centering	\lotop{$\varphi^{a_1\alpha_1}$}{$\varphi^{a_2\alpha_2}$}{black}
	\caption{Diagram that contribute for 2-point current. }
	\label{diags-bi-adjoint}
\end{figure}

\subsubsection*{Example}
The 2-point off-shell current in the regulated forward limit can be obtained from the trivalent diagrams shown in Fig.\ref{diags-bi-adjoint} and their permutations. The contributing diagrams coincide with the half-ladders that appear in the forward approach to thermal currents \cite{Brandt:2006aj}. Using momentum conservation and renaming the independent momentum by $k$, a simple calculation gives
\begin{align}
\ii \mathcal A^{\delta d, \alpha_1 a_1,\alpha_2 a_2, \epsilon e}
(p, \hbar  k ) &= \ii y^2\Bigg[ \left(
f^{\delta  \alpha _1 \alpha _3} f^{\epsilon  \alpha _2 \alpha _3} f^{c d a_1} f^{c e a_2}-f^{\delta  \alpha _2 \alpha _3} f^{\epsilon  \alpha _1 \alpha _3} f^{c d a_2} f^{c e a_1} 
\right)\frac{1}{\hbar p\cdot k}\\
&  +\frac 14 \left(f^{\delta  \alpha _2 \alpha _3} f^{\epsilon  \alpha _1 \alpha _3} f^{c d a_2} f^{c e a_1}+f^{\delta  \alpha _1 \alpha _3} f^{\epsilon  \alpha _2 \alpha _3} f^{c d a_1} f^{c e a_2}\right) \frac{k^2}{(p\cdot  k)^2}
\Bigg] +\mathcal O (\hbar), \nonumber
\end{align}
where the seemingly singular term in the first line vanishes upon computing the trace. Hence, calculating the trace  we obtain 
\begin{align}
\bar {\mathcal A}^{a_1 a_2, \alpha_1 \alpha_2}(p,k)= y^2C_2^2 \delta^{a_1a_2}\delta^{\alpha_1 \alpha_2} \frac 1 2 \frac{ k^2}{(p\cdot k)^2}.
\label{2-point-bi-adjoint}
\end{align} 
The squared retarded propagator has been left implicit and can be recovered by the analytic continuation  $k_0 \to k_0 +\ii o$, where $o$ is a small positive number. Using now Eq.\eqref{off-shell-current} and the integrals computed in Appendix \ref{integrals} (with $f_-(p_0)$) we find that in the 
the high temperature limit the 2-point current reads
\begin{align}
\Pi^{a_1 a_2, \alpha_1 \alpha_2}(k)
=\delta^{a_1a_2}\delta^{\alpha_1 \alpha_2}\frac{C_2^2 y^2}{2}\frac{T^2}{96 \pi}\left(\frac{k_0^2}{|\mathbf{k}|^2}-1\right)
\left[
-2+\frac{k_0}{|\bm k|}\log \left(\frac{|\bm k|+k_0 + \ii o}{-|\bm k|+k_0 + \ii o}\right)\right],
\end{align}
which agrees with cubic theory  in Ref.\cite{Brandt:2012cp}. In the limit where $k_0 \ll |\mathbf k|$ we have 
\begin{align}
\Pi^{a_1 a_2, \alpha_1 \alpha_2}(k)
\approx \delta^{a_1a_2}\delta^{\alpha_1 \alpha_2}\frac{C_2^2 y^2}{2}\frac{T^2}{96 \pi} \left(2-\ii \pi \frac{k_0}{|\bm k|}  \right),
\end{align}
where the imaginary part corresponds to Landau damping.

\subsection{Schwinger-Keldysh for bi-adjoints and kinetic theory}
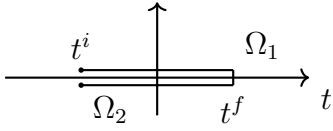
\begin{figure}
  	\centering
  	\begin{tikzpicture}[thick,decoration={%
	markings,%
	mark=at position 2.5cm with {\arrow[black]{stealth};}}]
\draw[->] (-2.,0) -- (2,0)  node[anchor=north west] {$t$};  
\draw[->] (0,-.5) -- (0,1) node[anchor=south east] {};
\draw(-1.,0.1) -- (1.,0.1)  node[anchor=south west ] {$\Omega_1$};
\draw(-1.,-0.1)node[anchor=north west] {$\Omega_2$} -- (1.,-0.1);
\draw(1.,0.1) -- (1.,-0.1)node[below ]{$t^f$};
\filldraw [black] (-1,0.1) circle [radius=.6pt]node[above ]{$t^i$} ;
\filldraw [black] (-1,-0.1) circle [radius=.6pt];    
\end{tikzpicture}
  	\caption{Representation of the Schwinger-Keldysh path. The small vertical line at $t_f$ does not contribute to the path integral. Phase-space variables in the upper contour and lower contour  are labeled as $w_1$ and $w_2$, respectively.}
  	\label{SK-contour}
  \end{figure}

We proceed now to develop a classical kinetic theory for bi-adjoint scalars along the lines of the Schwinger-Keldysh worldline approach by Mueller-Venugopalan \cite{Mueller:2019gjj}.   Suppose that we are interested in the description of the time evolution of the matrix density operator
\begin{align}
\rho (t^f)= U(t^f, t^i) \rho (t^i) U (t^i,t^f), 
\end{align}
where $U(t,t')$ is the evolution operator with $t^i$ and $t^f$ denoting the initial and final times, respectively.
Following the usual Schwinger-Keldysh approach \cite{Das:1997gg} we double the degrees of freedom and 
consider the difference 
\begin{equation}
\mathcal L = \mathcal L_{\text {BA},1} (\varphi_1, \partial \varphi_1 ) -\mathcal L_{\text {BA},2} (\varphi_2, \partial \varphi_2),
\end{equation}
where the Lagrangians $\mathcal L_{\text{BA},1} (\varphi_1, \partial \varphi_1 )$ and $\mathcal L_{\text{BA},2}(\varphi_2, \partial \varphi_2)$ are associated with the upper and lower branches of the contour shown in Fig.\ref{SK-contour}. Now,
using a quantum/background ($\varphi_i
\to\bar \varphi_i+\varphi_i$) expansion leads to a quadratic action of the form  
\begin{equation}
\begin{aligned}
\mathcal L^{(2)} = \frac 1 2 
\bar\varphi_1^{a\alpha} ( -\delta^{a\alpha}\delta^{b\beta}(\partial^2 + m^2)+y \hat \varphi_1^{a\alpha,b\beta} ) \bar\varphi_1^{b\beta}-
\frac 1 2 
\bar\varphi_2^{a\alpha} ( -\delta^{a\alpha}\delta^{b\beta}(\partial^2 + m^2)+y \hat \varphi_2^{a\alpha,b\beta} ) \bar\varphi_2^{b\beta},
\label{Schwinger-Keldysh-fields} 
\end{aligned}
\end{equation}
where  
$ \hat\varphi^{a\alpha,b\beta}_i:=
f^{abc}\tilde f^{\alpha\beta\gamma} \varphi^{c\gamma}_i.
$
Hence the particle matrix-valued Hamiltonian associated with each Lagrangian can be read-off from Eq.\eqref{Schwinger-Keldysh-fields}
\begin{align}
H_i= \frac 1 2  \left[ \delta^{a\alpha}\delta^{b\beta}(-p_i^2+m^2)-y\hat \varphi_i^{a\alpha,b\beta} \right].
\end{align} 
Notice that the total Lagrangian \eqref{Schwinger-Keldysh-fields} has a  diagonal matrix structure w.r.t.  to the  indices of $\varphi_i^{a\alpha}$ in addition to the matrix structure due to color indices. 

Its diagonal form implies that the path integral representation of the dressed propagator\footnote{Dressed propagators have been developed in a worldline representation for a variety of models, see Ref.\cite{Edwards:2019eby} and references therein for a review.} on the Schwinger-Keldysh contour is simply 
\begin{align}
G_\Omega:= G_\Omega\big(x_1^i,x_1^f; x_2^i, x_2^f;  \varphi_1,\varphi_2\big) =G_{1} (x_1^i, x_1^f; \varphi_1)
G_{2} (x_2^i, x_2^f; \varphi_2)\Big|_{x_1^f=x_2^f}\ ,
\end{align}
where phase space variables $w$ evaluated at initial and final times are written as  $w(t^i)=w^i$, $w(t^f)=w^f$, respectively. Here the variables $x_1$, $x_2$ describe  upper and lower contours with the  boundary condition $x_1^f=x_2^f$.   Inserting a worldline path integral representation for each dressed propagator then leads to 
\begin{align}
G_{\Omega}
= 
\int \frac{\mathcal D [e_1, e_2]}{\text{vol}(\text {Gauge})} 
\int \mathcal D [x_1,x_2] 
\int \mathcal D [p_1,p_2] 
\ \mathbf T  e^{\ii (S_1-S_2)}, 
\end{align}
where $\mathbf T$ denotes time ordering and   $\mathcal D[w_1,
w_2]:=\mathcal D w_1 \mathcal D w_2$.
Hence, after introducing complete sets of (initial) states, we find that the evolution of the matrix density operator $\rho^f:=\braket{x_1^f |\rho (t^f)| x_2^f}$ is given by  
\begin{align}
\rho^f_\Omega
:=\int\dd^4 x_1^i
\int\dd^4 x_2^i \ \rho (x_1^i, x_2^i) G_\Omega\,,
\label{def-density-final}
\end{align}
which is to be understood as the dressed propagator evaluated over the Schwinger-Keldysh contour $\Omega$ (see Fig.\ref{SK-contour}) weighted by the density matrix. Closing the contour leads to the real time effective action in Ref.~\cite{Mueller:2019gjj}. This definition is a specialization of the QFT case applied to worldlines, see e.g.,\cite{Gozzi:2010iu, Jeon:2004dh, Jeon:2013zga} . 

In the following we will
introduce the auxiliary variables $\psi^a$ and $\phi^\alpha$ to replace time ordering in path integrals following  Ref.\cite{Bastianelli:2021rbt}, where the interested reader can find details. 
The worldline action  in phase-space  for the bi-adjoint scalar in real time $\tau$ is  
\begin{align}
S_{\text{BA}} =\int \dd\tau 
\Big [ p_\mu \dot x^\mu + \ii \bar \psi_a \dot \psi^a  + \ii \bar \phi_\alpha \dot \phi^\alpha  
-e H - a J -  {\tilde a}{\tilde  J}  \Big ]  \;,
\label{action1}
\end{align}
where $H, J,\tilde J$ denote first class constraints 
\begin{align}
H &  = \frac12 ( -p^2+ m^2 +y\, C^a  \varphi^{a\alpha}(x) {\tilde C}^{\alpha} ) \;, \qquad 
C^a= -\ii f^{abc} \bar \psi ^b \psi^c \;,
\quad {\tilde C}^\alpha = -\ii \tilde f^{\alpha\beta\gamma} \bar \phi^\beta \phi^\gamma \;,
\label{ham}\\
J &  =  \bar \psi_a \psi^a -s \;, \qquad
\tilde J   = \bar \phi_{\alpha} \phi^\alpha -\tilde s    \;,
\label{cons2} 
\end{align}
which we identify with the particle Hamiltonian, the color charges and the particle current, respectively.
To avoid cluttered expressions we  define  $\mathsf P:=(p_\mu, \ii \bar \psi_a, \ii \bar \phi_\alpha )$, $\mathsf X:= (x^\mu, \psi^a, \phi^\alpha)$ and  group together the first class constraints  into $\mathsf e \mathsf H^t $, where $\mathsf e  := (e,  a , \tilde a)$ and
$\mathsf H:= (H, J, \tilde J)$.  Here $\mathsf A^t$ denotes transpose of $\mathsf A$\footnote{Bold  letters $\mathsf A$, $\mathsf B$,  denote  matrices.}. In condensed notation the worldline action is then 
\begin{align}
S_{\text{BA}}:=\int \dd\tau\ \left(\mathsf P \dot {\mathsf X}^t -\mathsf e \mathsf H^t\right).
\end{align}

 We may now introduce color variables that   
exchange time-ordering by an additional integration over auxiliary variables $\psi,\phi$. The dressed propagator then 
reads\footnote{In the path integral measures it is understood that only the variables in $\mathsf P$ are considering omitting factors of $\ii$.}
\begin{align}
G_\Omega[\mathsf X_1^i, \mathsf X_2^f;\varphi_1, \varphi_2] = 
\int \frac{\mathcal D [\mathsf e_1, \mathsf e _2]}{\text{vol}(\text {Gauge})} 
\int \mathcal D [\mathsf X_1,\mathsf X_2] 
\int \mathcal D [\mathsf P_1, \mathsf P_2] 
\ e^{\ii (S_{\text{BA},1}-S_{\text{BA},2})}\, .
\end{align}
Setting $\mathrm{e}=2T$, were $T$ is the so-called Schwinger proper time, produces the usual representation of the dressed propagator but the message of Ref.\cite{Mueller:2019gjj} is to keep the einbein unfixed to perform the path integral, which we will also do here. 
Hence the 
path integral representation of the effective action is
\begin{align}
\Gamma_\Omega[\varphi_1, \varphi_2]= \int\dd \mathsf X_1^i
\int\dd \mathsf X_2^i \ \rho (\mathsf X_1^i, \mathsf X_2^i) \int \frac{\mathcal D [\mathsf e_1, \mathsf e _2]}{\text{vol}(\text {Gauge})} 
\int \mathcal D [\mathsf X_1,\mathsf X_2] 
\int \mathcal D [\mathsf P_1, \mathsf P_2] 
\ e^{\ii (S_{\text{BA},1}-S_{\text{BA},2})}, 
\end{align}
where $\dd \mathsf X:= \dd^4 x\, \dd \psi \dd \bar \psi \,\dd \phi \dd\bar \phi$. 
Now we introduce 
Schwinger-Keldysh (SK) coordinates 
\begin{align}
z_R:= \frac{1}{2}(z_1+z_2), 
\qquad 
z_A:= z_1-z_2,
\end{align} 
for all phase-space coordinates. We interpret $z_A$ as a quantum degree of freedom measured in units of $\hbar$. We can then obtain the so-called  truncated Wigner approximation \cite{Polkovnikov:2009ys} by expanding in powers of $\hbar$   
\begin{align}
S_{\Omega, \text{BA}}=
\int \dd \tau \left(\mathsf P_R \dot {\mathsf X}_A^t +\mathsf P_A\dot {\mathsf X}_R^t -\mathsf e_R \mathsf H_A^t-\mathsf e_A \mathsf H_R^t\right), 
\end{align}
where we have kept  orders up to $\mathcal O (z_A)$ and we have defined the following quantities:
\begin{equation}
\begin{aligned}
\mathsf H_R=&\left(  \frac 12(-p_R^2+m^2+ y  
C_R^a \varphi_R^{a\alpha} \tilde C_R^\alpha),\quad J_R, \quad \tilde J_R\right),
\label{Hr-Wigner} \\
\mathsf H_A=&\left( ( - p_A \cdot p_R+y \frac 14 x_A^\mu  
C_R^a \partial_\mu\varphi_R^{a\alpha} 
\tilde C_R^\alpha), \quad \bar \psi_{Aa}   \psi_R^a+ \bar \psi_{Ra}   \psi_A^a,\quad
\bar \phi_{A\alpha}  \phi_R^\alpha+ \bar \phi_{R\alpha} \phi_A^\alpha \right).
\end{aligned} 
\end{equation}
Notice the absence of mixing terms involving $\bar \psi^b_R \psi^c_A$ or 
$\bar \phi^\beta_R \phi^\gamma_A$, which vanish due to the antisymmetry of the structure constants. Using the e.o.m we can rewrite the action as:
\begin{align}
S_{\Omega, \text{BA}}=
\mathsf P_R^i\cdot  {\mathsf X}_A^{i,t}-\mathsf e_A \mathsf H_R^t-\int \dd\tau \, \left(\dot{\mathsf P}_R+ \frac {\partial H}{ \partial \mathsf X_A^t}\right) \mathsf X_A^t+
\int \dd\tau\,  \mathsf P_A\left(\dot {\mathsf X}_R^t-\frac{\partial H}{ \partial {\mathsf P}_A}\right),
\end{align} 
where we notice that the boundary terms obtained through integration by parts  produce the 
Wigner function 
\begin{align}
W(\mathsf X^i_R,\mathsf P^i_R):
=
\int \dd \mathsf X_A^i  e^{\ii \mathsf  P_R^i \cdot {\mathsf X}_A^i}
\rho (\mathsf X_R^i+\frac 1 2\mathsf X_A^i, 
\mathsf X_R^i-\frac 1 2\mathsf X_A^i).
\end{align}
Finally, the Schwinger-Keldysh real-time effective action for the bi-adjoint scalars is 
\begin{equation}
\begin{aligned}
\Gamma_\Omega= \int &
\dd \mathsf X^i_R \int \dd \mathsf P^i_R \,
W(\mathsf X^i_R,\mathsf P^i_R) \\
&\times \int \mathcal D \mathsf e_R \mathcal  \int  \mathcal  D \mathsf X_R
\int 
\mathcal  D \mathsf P_R
\prod\limits_\tau 
\delta(P_R^2-m^2)
\delta (J_R) \delta (\tilde J_R)
\delta( \dot {\mathsf P}_R 
- \dot {\bar {\mathsf P}}
)
\delta(\dot {\mathsf X}_R-\dot {\bar {\mathsf X}}),
\end{aligned}
\end{equation}
where $\bar {\mathsf X}$ and $\bar {\mathsf P}$ satisfy classical equation of motion and $P_R^2$ can be read-off from Eq.\eqref{Hr-Wigner}.
Following \cite{Mueller:2019gjj},
the Liuoville's equation then reads
\begin{align}
\frac{\dd }{\dd \tau} W (\mathsf X, \mathsf P)= \left(\dot {x}^\mu \frac{\partial}{\partial x^\mu} +\dot {p}^\mu \frac{\partial}{\partial {  p}^\mu}+\dot { \psi}^a \frac{\partial}{\partial { \psi}^a}+
\dot { \bar \psi}^a \frac{\partial}{\partial {\bar  \psi}^a}+\dot {\phi}^a \frac{\partial}{\partial {\phi}^a}+
\dot {\bar \phi}^a \frac{\partial}{\partial {\bar  \phi}^a}\right)  W (\mathsf X, \mathsf P)=0
\end{align}
for some given initial condition $W(\mathsf X_R^i, \mathsf P_R^i)$. Let us mention that in 
the field theory case,
the appearance of the on-shell delta function is required to conserve momentum at each vertex 
\cite{Jeon:2004dh}. 
\subsection{Vlasov-type equation for bi-adjoint scalars}
For our purposes it will be more convenient to work directly with the classical limits of the color charges $C^a$ and $\tilde C^\alpha$ 
instead of the auxiliary ones. In the classical limit the charges defined in Eq.\eqref{ham}  correspond to classical color charges for large representations in a coherent state basis. These are controlled by $s$ and $\tilde s$ in \eqref{cons2} as can be seen by performing explicit path integration over auxiliary variables after gauge fixing (See appendix A of \cite{Ahmadiniaz:2015xoa}).   Therefore in the classical limit we may simply set
\begin{align}
C^a \to c^a, \qquad \tilde C^\alpha \to c^\alpha.
\label{Ctoc}
\end{align}
We can derive classical kinetic equations from the action \eqref{action1} setting $a=\tilde a=0$ and  $e=-1$ (so $p=\dot x$). We obtain
\begin{align}
-\dot p^\mu + \frac{ y }{2} c^a \tilde c ^\alpha \partial^\mu \varphi^{a \alpha}=0, \\  
\ii  \dot \psi ^a +  \frac y {2} \frac{\partial C ^{c}}{\partial \bar \psi _a} \varphi^{c \alpha}
\tilde C^\alpha=0, \\
\ii  \dot {\bar \psi} ^a -\frac y {2} \frac{\partial C ^{c}}{\partial \psi _a} \varphi^{c \alpha}
\tilde C^\alpha=0,
\end{align}
with an additional pair of equations for $\phi^\alpha$.   
Since $\dot C^a= -\ii f^{abc}(\dot{\bar \psi}^b \psi^c +\bar \psi^b \dot{\psi}^c)$, we may pack together the last two equations leading to
\begin{align}
\dot c^a=&   \frac y 2 f^{abc} c^b \tilde c^{\alpha}
\varphi^{c \alpha},
\end{align} 
where we have used the the Jacobi identity. The additional equation for $\tilde c^\alpha$ can be obtained similarly. 
Therefore the Liouville's equation
$\dd f/\dd\tau=0$ for the 
semi-classical distribution function $f(x, p, c, \tilde c)$ leads to the the Vlasov-type equation for bi-adjoint scalars
\begin{align}
\left(p^\mu \frac{\partial }{\partial x^\mu}+ \frac 12 
y c^a \tilde c ^\alpha \partial^\mu \varphi^{a \alpha}
\frac{\partial }{\partial p^\mu}
+ \frac 12 y f^{abc}  c^b \tilde c^{\alpha}
\varphi^{c \alpha} \frac{\partial }{\partial  c^a}
+\frac 12 y f^{\alpha \beta \gamma } \tilde c^\beta c^{a}
\varphi^{ a \gamma} \frac{\partial }{\partial \tilde c^{\alpha}}\right)f=0.
\label{Vlasov-bi-adjoint}
\end{align}
Perturbative thermal currents can be obtained from 
\begin{align}
J^{a \alpha}(x)=y \int \dd \Phi (p)\int   
\dd c  \int \dd \tilde c \ c^a \tilde c^\alpha
f(x, p, c, \tilde c),
\label{Vlasov-current}
\end{align}
where the color phase space invariant measure is defined in Eq.\eqref{DIPSC}.

\subsubsection*{Example}
Let us  check the consistency of the semi-classical kinetic theory just constructed for  the 2-point thermal current. Expanding the distribution function around equilibrium we have
\begin{align}
f(x,p, c, \tilde c)= f^{(0)}(p_0)+ y f^{(1)}(x,p,c,\tilde c)
+ \dots,
\label{expansion-bi-adjoint}
\end{align} 
where $f^{(0)}(p_0)=1/(e^{\beta p_0}-1)$ is the Bose-Einstein distribution function and $\beta=1/T$. Moving to Fourier space and
plugging in Eq.\eqref{expansion-bi-adjoint} into \eqref{Vlasov-bi-adjoint} we find  
\begin{align}
\tilde f^{(1)}(k,p, c, \tilde c )=- c^a \tilde c^\alpha
\frac{k^\mu}{k\cdot p}
\frac{\partial f^{(0)}}{\partial p^\mu} \tilde \varphi^{a \alpha} (k),
\end{align}
where $\tilde f^{(1)}(k,p, c, \tilde c )$ denotes the Fourier transform of $f(x,p, c, \tilde c)$. Inserting this equation into the definition of the current  we obtain
\begin{align}
\tilde J^{a\alpha}= \frac{y^2}{2}  \int \dd \Phi (p)  \int
\dd c \int \dd \tilde c \ c^a \tilde c ^\alpha
c^b \tilde c ^\beta 
\ f^{(0)} (p_0) \frac{k^2}{ (k\cdot p)^2} \varphi^{b \beta },
\label{current-LO}
\end{align}
which leads to 
\begin{align}
\Pi^{ab, \alpha\beta}(k)=\delta^{ab} \delta^{\alpha\beta}\frac{ C_2^2 y^2}{2}  \int \dd \Phi (p) 
\  f^{(0)} (p_0) \frac{k^2}{ (k\cdot p)^2},
\end{align}
where we have used the identities in Eq.\eqref{identities-phase-space}. This is reproduces exactly the result we obtained from QFT  in Eq.\eqref{2-point-bi-adjoint}.

\section{Spin}
\label{gauge-spin}
Classical color charges  and  spin vectors have in common that they may be associated with expectation values of certain operators  with respect to coherent states. Indeed, provided one is able to build coherent states that furnish an irreducible representation of the Lie group, say $SU(N)$, then
the following properties   hold \cite{Yaffe:1981vf} 
\begin{align}
\braket{\psi|\mathbb A|\psi}&= \text{finite},\\
\braket{\psi|\mathbb A \mathbb B|\psi}&= \braket{\psi|\mathbb A|\psi} \braket{\psi|\mathbb B|\psi} + \dots \,,
\end{align}
where the ellipsis means terms that do not contribute in the classical limit.
In particular, one may use the Schwinger-boson formalism for $SU(N)$ to build
explicit realizations
of these states and  show the above properties. For
color this construction was used in Ref.\cite{delaCruz:2020bbn} to describe colored observables  while for spin in Ref.~\cite{Aoude:2021oqj}.
Therefore one expects that the dynamics of color charges and spin share  some similarities as studied in Refs.~\cite{delaCruz:2021gjp,Maybee:2019jus, Cristofoli:2021jas}.

The description of classical spin may also be done by 
introducing (integer) higher-spin massive particles  described by symmetric traceless rank-$s$ tensor fields $\varphi_s^{a_1\cdots a_s}$ \cite{Singh:1974qz}, where for brevity we will suppress its indices henceforth. Lorentz generators also carry (symmetrized) sets of indices, which we will also suppress  but use  matrix notation 
$\mathsf M^{\mu\nu}$ to indicate that the contraction of indices is  understood as matrix multiplication. 
Let $\varepsilon(p):=\varepsilon (s,p)$ be the   polarization tensors of the massive particle with momentum $p$. Denoting by $\varepsilon (\tilde p)\cdot 
\varepsilon (p)$ the contraction of the tensor indices of the polarization tensors, the relation between classical spin tensors and Lorentz generators $\mathsf M^{\mu\nu}$ is given by  the identities
\begin{equation}
\begin{aligned}
\varepsilon (\tilde p)
\mathsf M^{\mu\nu} \varepsilon ( p)=&
S^{\mu\nu}
\varepsilon (\tilde p)\cdot 
\varepsilon (p) +\dots,\\
\varepsilon (\tilde p) \{ \mathsf M^{\mu\nu}, \mathsf M^{\rho \sigma}\}  \varepsilon ( p)&=S^{\mu\nu}S^{\rho\sigma}\varepsilon (\tilde p)\cdot 
\varepsilon (p) + \dots,
\end{aligned}
\label{spin-identities}
\end{equation}
where $\{\mathsf A, \mathsf B\}:= \frac 12(\mathsf A \mathsf B+\mathsf A \mathsf B)$ and $\tilde p:=p-q$. We refer the interested reader to Ref.\cite{Bern:2020buy}  for details on the spin formalism.
The ellipsis denote terms that do not contribute in the classical limit. The spin vector and spin tensor are related through Eq.\eqref{spin-tensor-to-vector}. The spin vector is constrained by the so-called covariant spin supplementary condition 
\begin{align}
p_\mu S^{\mu\nu}=0.
\label{spin-supplementary-condition}
\end{align}
\begin{figure}[tb]
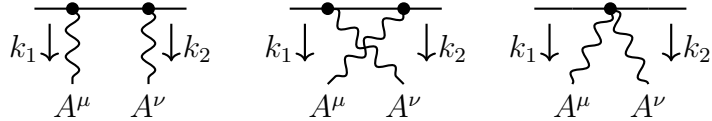

	\centering	\lotop{$A^\mu$}{$A^\nu$}{black}
	\lotopB{$A^\nu$}{$A^\mu$}{black}
	\lotopC{$A^\mu$}{$A^\nu$}{black}
		\caption{2-point graphs that contribute to the current. Besides the usual interactions in scalar QED, the higher-spin model allows interactions proportional to Lorenz generators. }
	\label{topos-spin}
\end{figure}
\subsection{Spinning off-shell currents}

Let $F^{\mu\nu}$ be the usual Maxwell field strength and $D^\mu$ the corresponding covariant derivative. The Lagrangian density of the higher-spin scalar electrodynamics is given by 
\cite{Bern:2020buy} 
\begin{align}
\mathcal L_{\text{EM}} =-\frac 1 4 F^{\mu\nu} F_{\mu\nu}+ D^\dagger_\mu \bar \varphi_s D^\mu \varphi_s-m^2\bar \varphi_s \varphi_s+
e  F_{\mu\nu} \bar \varphi_s \mathsf M^{\mu\nu}\varphi_s,
\label{lagrangian-spin} 
\end{align}
where $\varphi_s$ are higher-spin fields.
Using the double copy, this theory has been used recently in the context of scattering amplitudes to describe the Post-Minkowskian two-body problem of spinning particles in gravity \cite{Kosmopoulos:2021zoq, Bern:2022kto}. 
Classical currents including spin contributions may be  computed from Eq.\eqref{off-shell-current} with $\widehat {\text{Tr}}(\bullet)$ now defined as
\begin{align}
\widehat{\text{Tr}}(\bullet):= \text{Tr}_s (\bullet),
\end{align}
where by construction satisfies the same properties of phase-space integration over classical spin given in Eq.\eqref{identities-phase-space}. E.g., $\text{Tr}_0(\bullet)=\text{Id} (\bullet)$ is equivalent to $\int \dd s=1 $ classically. 
We can now use Eq.\eqref{off-shell-current} to perform calculations. Notice that in the forward limit we may set  $\varepsilon (p) \cdot \varepsilon (p)=1$.

For instance, consider the 2-point current (See Fig.\ref{topos-spin}). It receives contributions from the same diagrams as in scalar QED but has, in addition, contributions  coming from the last term in Eq.\eqref{lagrangian-spin}. 
The off-shell current has the structure
\begin{align}
\mathcal A^{\mu\nu}(p,k)=\varepsilon_s (p)\cdot   A^{\mu\nu} (p,k) \cdot \varepsilon_s (p)\,.
\end{align}
However, from the properties of the trace (or equivalently from the phase integration) we see that
only those  terms appearing quadratically may lead to meaningful contributions. Keeping this in mind,
the current now can be written as follows:
\begin{align}
A^{\mu\nu}(p,k)= A^{\mu\nu}_{0} (p, k) + A _{s}^{\mu\nu}(p, k) \, , 
\label{def-current}
\end{align}
where the scalar part  is the current in scalar QED
\begin{align}
\ii A^{\mu\nu}_{0} (p, k)= 2 \ii e^2 \left[ \eta^{\mu\nu} -\frac{\left(k^{\mu }+2 p^{\mu }\right) \left(k^{\nu }+2 p^{\nu }\right)}{2 \left(2 k\cdot p+k^2\right)}-\frac{\left(k^{\mu }-2 p^{\mu }\right) \left(k^{\nu }-2 p^{\nu }\right)}{2 \left(k^2-2 k\cdot p\right)}\right],
\end{align}
which in the classical limit leads to
\begin{align}
A_0^{\mu \nu}(k,p):= 2e^2\overline {\Pi}^{\mu\nu}_0= 2e^2\left( \eta^{\mu\nu}-\frac{p^\mu k^\nu+p^\nu k^\mu}{k\cdot p}+\frac{k^2 p^\mu p^\nu}{(k\cdot p)^2}\right).
\label{current-QED-spin}
\end{align}
On the other hand, the spin-dependent contribution reads
\begin{equation}
\begin{aligned}
\ii A _{s}^{\mu\nu}(p, k)=-2\ii e^2\Bigg\{ \left[
\frac{(2p+k)^\mu}{2p\cdot k+k^2}+
\frac{(2p-k)^\mu}{-2p\cdot k+k^2}\right]
\ii (k\cdot \mathsf M)^\nu-\mu\leftrightarrow\nu
\\
+2k_\alpha k_\beta  \left(
\frac{\mathsf  M^{\alpha\mu}\mathsf M^{\beta\nu}}{2p\cdot k+k^2}+
\frac{\mathsf M^{\beta\nu}\mathsf M^{\alpha\mu}}{-2p\cdot k+k^2}\right)
\Bigg\},
\end{aligned}
\end{equation}
where $k_\mu \mathsf M^{\mu\nu}:=(k\cdot \mathsf M)^\nu$.
Now, keeping only contributions quadratic in spin and using Eq.\eqref{spin-identities}, we find that in the classical limit\footnote{In KMOC the spin vectors are of $\mathcal O(\hbar)$ and so spin-squared terms appear as terms of $\mathcal O (\hbar^2)$. The classical result shown here  it is understood as the expectation value w.r.t coherent states \cite{Aoude:2021oqj}.}
\begin{align}
A _{s}^{\mu\nu}(p, k)|_{S^2}:=2 e^2 \overline \Pi_s^{\mu\nu}=2  e^2 \frac{k^2 }{(p\cdot k)^2}(k \cdot S)^\mu
(k \cdot S)^\nu,
\end{align}
which satisfies the Ward identity $k_\mu A_s^{\mu\nu} (p, k)|_{S^2}=0$. To reach this form we have also used the usual algebra  of $\mathsf M^{\mu\nu}$.
We will see in Section \ref{classical-comps} that this matches the result computed from a classical  perspective.  Computing the trace, or equivalently the phase space integration, we obtain
\begin{align}
\text{Tr}_s( A_s^{\mu\nu}(p,k))=-\frac{4 e^2 \mathfrak s^2}{3}k^2 \left( \frac{1}{m^2}\overline{\Pi}^{\mu \nu}_0+
\frac{k^2\eta^{\mu\nu}-k^\mu k^\nu}{(k\cdot p)^2}\right).
\end{align}
Therefore the off-shell current in the classical limit including spin contributions is
\begin{align}
\bar {\mathcal A}^{\mu\nu} (p,k)
=2 e^2 \left[\left( 1-\frac{2\mathfrak s^2 k^2}{3m^2}\right)\overline{\Pi}_0^{\mu\nu}-  \left( \frac{2\mathfrak s^2 k^2}{3}\right) \frac{\overline{N}^{\mu\nu}}{(k\cdot p)^2}
\right], 
\label{res-spin-2-point}
\end{align} 
where $\overline{ N}^{\mu\nu}:=k^2 \eta^{\mu\nu}-k^\mu k^\nu$. 

A few comments are in order. Recall that we still need to be integrate over phase space. Notice that the numerator $\overline{N}^{\mu\nu}$ in  second term in parenthesis also appears in the renormalization of QED. 
Let's perform  the analytic continuation $k_0 \to k_0+\ii o$  as usual and  consider the high temperature regime. Since $\overline{N}^{\mu\nu}$ is independent of the momentum $p$ we require the integral
\begin{align}
\int \dd \Phi(p)\ \frac{f_{+}^{(0)}(p_0)}{(k\cdot p)^2}=-\dfrac{1}{16\pi^2 k^2} \left[\dfrac{1}{\epsilon}+\dfrac{k_0}{|\bm k|} \log \left(\frac{|\bm k|+k_0+\ii o}{-|\bm k|+k_0+\ii o} \right)+\log(\beta^2 \mu^2)\right]+\mathcal{O}(\epsilon), 
\end{align}
which has been evaluated in $d=4-2\epsilon$ (see Appendix \ref{integrals}). The temperature-independent divergence cancels after subtracting the zero temperature contribution while the temperature dependent log  may be combined with the $\log (m^2/\mu^2)$  in the renormalization of $\Pi^{\mu\nu}$. Notice also that Eq.\eqref{res-spin-2-point} is strictly valid only where $m\ne 0$. The reason is that the spin supplementary condition $p_\mu S^{\mu\nu}=0$ can no longer be used as a condition that fixes $S^{\mu\nu}$ uniquely as the spin tensor in the rest frame of the particle since there is no such frame for massless particles. Instead, for massless particles  one can choose a frame characterized by a
time-like vector $u^\mu$  and set $p^\mu \to u^\mu$
in Eq.\eqref{spin-supplementary-condition}  \cite{Weickgenannt:2019dks, Chen:2015gta}.  
\subsection{Comparison with semi-classical kinetic theory}
\label{spin}
\label{classical-comps}
In order to check the validity of our approach let us now consider a  classical perspective.
The generic form of the collisionless relativistic Boltzmann-Vlasov equation reads
\cite{Heinz:1984my, Israel:1978up}
%
\begin{align}
p^\mu \frac{\partial f}{\partial x^\mu}
+e \frac{\partial ({F}^\mu f)}{\partial p^\mu}=0,\quad
F^\mu=
F^{\mu\nu}p_\nu+ \frac 1 2\frac{\partial F^{\nu\rho}}{
	\partial x_\mu} S_{\nu\rho} \,.
\label{vlasov-spin-1}
\end{align}
This equation can be derived within Wigner function formalism applied for spin $\frac 12$ particles \cite{Weickgenannt:2020aaf, Gao:2019znl}.
Now, as in the spin-less case, let us perturb  the distribution function $f$ around equilibrium
\begin{align}
f= f^{(0)}(p_0)+e f^{(1)} (x,p, S)+
e^2 f^{(2)}(x,p, S)+\dots \, , 
\end{align}
where $f^{(0)} (p_0)$ is the Fermi-Dirac distribution function. We are interested in computing the associated current
\begin{align}
J^\mu(x)= e 
\int \dd \Phi(p) \int \dd s\  \left(p^\mu +S^{\mu\nu}\partial_\nu\right) f(x, p, S) \,.
\label{current-spin-QED}
\end{align}
Solving for the coupled system of Eq.\eqref{vlasov-spin-1} and Eq.\eqref{current-spin-QED} in momentum space, we obtain  
\begin{align}
\tilde f^{(1)}(k,p,S)=
\frac{\ii }{k\cdot p}\left[\frac{ k^\mu}{2} (k^\nu \tilde A ^\rho-k^\rho \tilde A^\nu ) \frac{\partial ( f^{(0)} S_{\nu\rho}) }{\partial p^\mu }+\ii p_\nu (k^\mu \tilde A^\nu-k^\nu \tilde A^\mu)\frac{\partial  f^{(0)} }{\partial p^\mu }
\right].
\end{align}
Plugging in this into the Eq.\eqref{current-spin-QED} and using integration by parts we can bring the current into the form
\begin{align}
\tilde J^{\mu,(1)} (k)= -e^2 \int \dd \Phi(p)\int \dd s \, \tilde{f}^{(0)} (p_0) \left[\overline{\Pi}^{\mu \nu}_0(k,p)+\overline{\Pi}_s^{\mu \nu}(k,p)\right]\tilde A_{\nu} \,,
\end{align}
where the spin-dependent contribution is
\begin{align}
\overline{\Pi}^{\mu \nu}_s(k,p)=&\frac{k^2}{(k\cdot p)^2} (k\cdot S)^\mu(k\cdot S)^\nu,  
\end{align}
and $\overline \Pi_0^{\mu \nu}(k,p)$ is given in Eq.\eqref{current-QED-spin}. This matches our results obtained using classical limits of off-shell currents.  
\section{Conclusions}
\label{conclusions}
We have extended the scattering amplitudes approach of Ref.\cite{delaCruz:2020cpc} in two ways: we have considered thermal currents for bi-adjoint particles and  spin.  To test its  validity, we have compared against computations based on iterative solutions of classical kinetic equations finding  agreement. 
The semi-classical kinetic equations for bi-adjoints were derived from the worldline approach in Sec.\ref{Bi-adjoint} following   Mueller-Venugopalan \cite{Mueller:2019gjj}, who have shown that in the classical limit kinetic equations follow from the Schwinger-Keldysh effective action adapted to worldlines. For the case of spin we have used the well-known Boltzmann-Vlasov equations for spinning particles. On the amplitudes side,  we have modeled classical spin through a higher-spin generalization of scalar QED.

Hard thermal loop actions can 
be computed from other methods including the worldline approach as found long time ago \cite{Venugopalan:2001hp,Jalilian-Marian:1999uob}. Moreover, 
the worldline setting of Ref.\cite{Mueller:2019gjj} suggests that these  may  also be derived from the recently proposed Worldline QFT approach \cite{Mogull:2020sak,Jakobsen:2021smu,Jakobsen:2021lvp, Jakobsen:2021zvh,Shi:2021qsb, Jakobsen:2022psy}. Therefore it would be interesting to investigate whether both thermal and classical observables  can be  unified in a worldline framework. 
Having introduced spin into the formalism of  Ref.\cite{delaCruz:2020cpc} the remaining task is to include collision functions. These have been studied in the context of classical transport in Ref.\cite{Litim:1999ns, Litim:1999id, Litim:1999ca}. We leave this for future work.

\addsec{Acknowledgements}
 We thank financial support from the Open Physics Hub at the Physics and Astronomy Department in Bologna. Some of the 
calculations in this paper were done with Feyncalc \cite{Mertig:1990an, Shtabovenko:2016sxi, Shtabovenko:2020gxv}.
We thank Fiorenzo Bastianelli and Francesco Comberiati for discussions and for collaboration on related projects.
We thank Niklas Mueller for clarifications regarding Ref.\cite{Mueller:2019gjj}. We also thank Andr\'es Luna and Maria Elena Tejeda-Yeomans for comments on a previous version of this manuscript.  
\appendix
\section{Massive phase-space integrals}
\label{integrals}
We can parametrize the momentum $p$ as $p=(p_0,|\bm p| \cos \theta, |\bm p|\sin \theta \bm1_{d-2})$,  where $\bm 1_{d-2}$ is a unit vector in $d-2$ dimensions. Hence   
recalling that $\dd \Phi(p):=\dd^d p /(2\pi)^{(d-1)}\,  \delta(p^2-m^2) \Theta(p_0)$
we may write down the measure as \cite{Abreu:2014cla}:
\begin{align}
\dd\Phi(p)= \frac{1}{(2\pi)^{d-1}}\dd \Omega_{d-2} \dd p_0 \dd |\bm p| \dd (\cos\theta) \ |\bm p|^{d-2} \Theta(p_0)\delta(p_0^2-|\bm p|^2-m^2)\sin^{d-4} \theta  ,
\end{align}
where 
\begin{equation}
\int \dd \Omega_{n}= \frac{2\pi^{n/2}}{\Gamma (n/2)}.
\end{equation}
We are the interested in solving the family of integrals
\begin{align}
I_{\pm}[a]:=&\int \dd \Phi(p)\ \frac{ f_{\pm}^{(0)}(p_0)}{(k\cdot p)^a}, 
\end{align}
where  $f_\pm^{(0)}(p_0)=1/(e^{\beta p_0}\pm1)$ and $\beta=1/T$. The simplest case is the case where $a=0$. Performing a change of variables $x=\beta p_0$ the result is proportional to the integral of
\begin{align}
\int_0^\infty \dd x\ (x^2-\beta^2 m^2)^{(d-3)/2} f_\pm\left(x/\beta\right)
\label{inte-zero} 
\end{align}
so in the high temperature limit, we obtain
\begin{align}
I_-[0]=\frac{T^{d-2}}{4\pi^{d/2}}\Gamma\left[d/ 2-1\right]\text{Li}_{d-2}(1),
\end{align}
where $\text{Li}_{n}(1)$ is the polylogarithm, which reduces to the Riemann zeta function  $\zeta(d-2)$. We also have   $I_+=(1-2^{3-d})I_-[0]$. The general case can be treated as follows. Without loss of generality we may choose $k=(k_0, |\bm k|, \bm 0_{d-2})$ and perform the analytic continuation  $k_0 \to k_0+\ii o$, hence 
$k\cdot p=k_0 p_0- |\bm p| |\bm k| \cos\theta + \ii o p_0$. Then, introducing the change of variables $\alpha = |\bm p|/p_0$ we have
\begin{align}
k\cdot p= p_0(k_0+\ii o- \alpha |\bm k| \cos \theta)
\end{align}
and so we may perform the angular integral  leading to 
\begin{align}
\int_{-1}^1 \dd( \cos\theta)  \frac{\sin^{d-4}\theta}{(k_0+ \ii o -\alpha |\bm k| \cos \theta)^a}=\frac{\sqrt{\pi} \, \Gamma\left[d/2-1\right]}{\Gamma[d/2-1/2](k_0+ \alpha \bm |k|)^a}\
{}_2F_1(a, (d-2)/2, d-2,B(\alpha)),
\end{align}
where  $d >2$ and ${}_2F_1(a, b, c,x)$ is the Gau\ss$\,$   hypergeometric function. Its argument is 
\begin{align}
B(\alpha):=2 \frac{\alpha |\bm k|}{k_0+\ii o+ \alpha |\bm k|}.
\end{align}
Now solving the integral over $\alpha$ by using the Dirac-delta and a change of variables $x=\beta p_0$,
we find 
\begin{align}
&I_\pm[a]= \beta^{2+a-d} \frac{(4\pi)^{(1-d)/2}}{\Gamma[(d-1)/2]}\\
&\times\int_0^\infty \dd x \  f_\pm(p_0/\beta) x^{-a} \frac{ \left(x^2-\beta^2 m^2\right)^{(d-3)/2}}
{(k_0+\sqrt{y}|\bm k|)^a} \, {}_2F_1(a, (d-2)/2, d-2,B(\sqrt{y}))\Big|_{y=(x^2-\beta^2 m^2)/x^2}\nonumber\, ,
\end{align}
which is  analytically regularized with $d=n-2\epsilon$ for $n>2$.  This integral is in general hard to evaluate analytically.
However, we can use this representation to obtain an expansion in powers of $\lambda=\beta m$ and take the leading order. It is easy to check that when $a=0$ this integral reduces to Eq.\eqref{inte-zero}. Other cases can be obtained similarly. For instance, the integral is finite for $d=4$ and $a=1$, so using
\begin{align}
{}_2F_1(1, 1, 2,B(\sqrt{y}))\Big|_{y=(x^2-\beta^2 m^2)/x^2}= \frac 12 \frac{|\bm k|+k_0}{ |\bm k|}
\log \left(\frac{|\bm k|+k_0+\ii o}{-|\bm k|+k_0+\ii o} \right) +\mathcal O(\lambda^2),
\end{align}
one obtains  standard results in the literature \cite{Laine:2016hma}. For $a=2$ we consider two cases in the main text (This integral has been studied  e.g., in Ref.\cite{Ayala:2008dn} using Mellin-Barnes techniques in the limit $|\bm k| \to 0$. Our results are in agreement). They are  
\begin{align}
I_+[2]=\begin{cases}
\dfrac{1}{96\pi|\bm k|^2\beta^2}\left[-C_1+
\dfrac{k_0}{C_2|\bm k|} \log \left(\frac{|\bm k|+k_0+\ii o}{-|\bm k|+k_0+\ii o} \right)\right]+\mathcal O (\epsilon),& n=6\\
\\
-\dfrac{1}{16\pi^2 k^2} \left[\dfrac{1}{\epsilon}+\dfrac{k_0}{|\bm k|} \log \left(\frac{|\bm k|+k_0+\ii o}{-|\bm k|+k_0+\ii o} \right)+\log(\beta^2 \mu^2)\right]+\mathcal{O}(\epsilon),& n=4 ,
\end{cases}
\end{align}  
where $\mu$ is some renormalization scale. Here $C_2=2$ and $C_1=1$ for $f_+(p_0)$ and $C_2 \leftrightarrow C_1$ for $ f_{-}(p_0)$. \section{3-point example}
\label{3-point-bi-adjoint}

\begin{figure}[t]
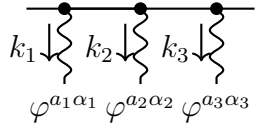

	\centering
	\nlotop{$\varphi^{a_1\alpha_1}$}{$\varphi^{a_2\alpha_2}$}{$\varphi^{a_3\alpha_3}$}
	\caption{3-point Feynman diagram. The remaining ones are obtained by permutations.}
	\label{diags-bi-adjoint-3}
\end{figure}
We can compute the 3-point current in a similar fashion considering all permutations of the diagrams in Fig.\ref{diags-bi-adjoint-3} leading to  
\begin{multline}
 \bar {\mathcal A}^{a_1\alpha_1, a_2\alpha_2, a_3\alpha_3}(p, k) = C_2^2 \frac{ y^3}{32} f^{a_1 a_2 a_3}
f^{\alpha_1 \alpha_2 \alpha_3} \\
\times
 \sum\limits_{\sigma \in \text{Cyclic}}\left(\frac{k_{\sigma_1}^2 k_{\sigma_2}^2}{(p\cdot   k_{\sigma_1})^2
	(p\cdot k_{\sigma_2})^2}+\frac{k_{\sigma_1}^2 k_{\sigma_1}^2}{(p\cdot k_{\sigma_1})^2 (p\cdot k_{\sigma_2})(p\cdot k_{\sigma_3})}\right),
\end{multline}
where $\text{Cyclic}$ is the set of cyclic permutations of $\{1,2,3\}$. 
This matches the same kinematics of the simple cubic theory as can be easily checked.  The classical result based on Eq.\eqref{Vlasov-bi-adjoint} can be obtained by an additional iteration of $f(x,p,c, \tilde c)$.
As usual,  there are  seemingly singular terms in the classical limit which vanish after computing the trace. 
Using this result we may implement a "classical double copy" replacement \cite{Goldberger:2016iau} with $y \to g $ and
\begin{align}
f^{\alpha_1\alpha_2 \alpha_3} \to \left[
\eta^{\mu_1\mu_2}(k_1-k_2)^{\mu_3}+
\eta^{\mu_2\mu_3}(k_2-k_3)^{\mu_1}+
\eta^{\mu_3\mu_1}(k_3-k_1)^{\mu_2} \right].
\end{align} 
Upon matching conventions this recovers  an Ansatz for the  logarithmic dependence in temperature $T$ of the 3-point function in QCD proposed in Ref. \cite{Brandt:1997rz}.

\bibliographystyle{JHEP}

\renewcommand\bibname{References} 
\ifdefined\phantomsection		
\phantomsection  
\else
\fi
\addcontentsline{toc}{section}{References}
\providecommand{\href}[2]{#2}\begingroup\raggedright\endgroup


\end{document}